\documentclass[aps,pra,twocolumn,showpacs,superscriptaddress]{revtex4-1}
\usepackage{amssymb}
\usepackage{amsmath}
\usepackage{amsthm}
\usepackage{txfonts}
\usepackage{graphicx}

\begin{document}

\title{Enhancement of quantum correlations between two particles under decoherence in finite temperature environment}

\emph{}
\author{Nasibollah Doustimotlagh}
\affiliation{State Key Laboratory of Low-Dimensional Quantum Physics and Department of Physics, Tsinghua University, Beijing 100084, China}
\author{Shuhao Wang}
\affiliation{State Key Laboratory of Low-Dimensional Quantum Physics and Department of Physics, Tsinghua University, Beijing 100084, China}
\author{Chenglong You}
\affiliation{Department of Physics \& Astronomy, Hearne Institute for Theoretical Physics, Louisiana State University, 202 Nicholson Hall, Baton Rouge, Louisiana 70803, USA}
\author{Gui-Lu Long}
\email{gllong@tsinghua.edu.cn}
\affiliation{State Key Laboratory of Low-Dimensional Quantum Physics and Department of Physics, Tsinghua University, Beijing 100084, China}
\affiliation{Tsinghua National Laboratory for Information Science
and Technology, Beijing 100084, China}
\affiliation{Collaborative Innovation Center of Quantum Matter, Beijing 100084, China}
\date{\today }

\begin{abstract}
Enhancing the quantum correlations in realistic quantum systems interacting with the environment of finite temperature is an important subject in quantum information processing. In this paper, we use weak measurement and measurement reversal to enhance the quantum correlations in a quantum system consisting of two particles. The transitions of the quantum correlations measured by the local quantum uncertainty of qubit-qubit and qutrit-qutrit quantum systems under generalized amplitude damping channels are investigated. We show that, after the weak measurement and measurement reversal, the joint system shows more robustness against decoherence.
\end{abstract}


\maketitle

\section{Introduction}
\label{intro}

Decoherence in realistic quantum systems affects quantum features in quantum information processing (QIP) severely \cite{book,yu2004}. Thus protecting quantum states under decoherence is an important subject in QIP tasks. Many schemes have been put forward to achieve this purpose, including dynamical decoupling \cite{viola1998,viola1999,zanardi1999}, decoherence free subspaces \cite{lidar1998,xu2012,feng2013}, quantum error correction code \cite{calderbank1996,steane1996,knill1997}, environment-assisted error correction scheme \cite{yu2013,yu2014}, quantum Zeno dynamics \cite{facchi2004,paz2012}, etc. A novel idea for protecting quantum states by weak measurement and measurement reversal has been proposed theoretically \cite{korotkov2010,cheong2012}, and it has been experimentally implemented in the last few years \cite{katz2008,kim2009,kim2012}. The researches focus on the fidelity and quantum entanglement of a quantum system protected by weak measurement and measurement reversal under decoherence \cite{sun2010,man2012,xiao2013,wang2014}.

It is widely believed that quantum entanglement is only one of the ingredients of quantum features \cite{horodecki2009}. As a larger family, quantum correlations are believed to reflect more about the quantumness in QIP \cite{modi2012}. Explicitly, quantum entanglement is a subset of quantum correlations for mixed states. In most QIP tasks, we always face the situation that the quantum system is a mixed state, especially when the quantum system suffers from decoherence. Therefore, it is desirable to study, and to protect, the quantum correlations in the realistic quantum systems under decoherence.
There are many kinds of quantifiers of quantum correlations, we adopt the local quantum uncertainty \cite{girolami2013,wang2013} for its  operability. 

\begin{figure}[!htp]
\begin{centering}
\includegraphics[width=7cm]{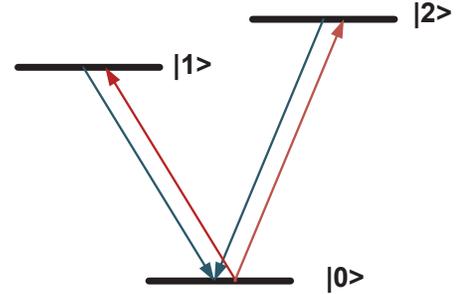}
\caption{(Color online) Energy level of a $V$-configuration particle. The transitions characterized by red lines only happen when the temperature is non-zero.}
\label{fig1}
\end{centering}
\end{figure}

We study the enhancement of quantum correlations for qubit-qubit and qutrit-qutrit quantum systems. It need to be noted that in the three-dimensional case, we suppose each of the two particles has $V$-configuration energy levels, as illustrated in Fig. \ref{fig1}. The extension to $\Lambda$-configuration can be naturally done by our approach. In this case, only the transitions from $|2\rangle$ and $|1\rangle$ to $|0\rangle$ are allowed, which simplifies our analysis. In order to characterize decoherence in a finite temperature environment, we use the generalized amplitude damping channel.

In this paper, we study the enhancement of quantum correlations by weak measurement and measurement reversal. We show that under decoherence, the quantum correlations between two particles can be enhanced. The remainder of this paper is organized as follows: In Sec. \ref{basic}, we give the preliminaries needed in the following parts. We will introduce the local quantum uncertainty and its closed form. The Kraus operators of the generalized amplitude damping for two- and three-dimensional quantum states having $V$-configuration energy levels are given. We have also shown the mathematical expressions for the weak measurement and measurement reversal operators. In Sec. \ref{enhance}, we investigate the enhancement of quantum correlations using the weak measurement and measurement reversal for the qubit-qubit Bell state with white noise, a non-symmetrical qubit-qubit mixed state, and the qutrit-qutrit Bell state with white noise. We have shown that the approach can be used to enhance the quantum correlations under decoherence. In Sec. \ref{con}, we have discussed the fidelity of the final output state, and give some conclusions.

\section{Basic theory}
\label{basic}

\subsection{Local quantum uncertainty}
The local quantum uncertainty (LQU) is defined as
\begin{equation}
\mathcal{U}_A=\min_{K^A} I(\rho_{AB}, K^A),
\end{equation}
where we have denoted the two particles as $A$ and $B$, the minimum is optimized over all the non-degenerate operators on A: $K^A = \Lambda^A \otimes \mathbb{I}_B$, and
\begin{equation}
I(\rho, K)= -\frac{1}{2}\text{Tr}\{[\sqrt{\rho},K^A]^2\}
\end{equation}
is the skew information \cite{wigner1963}. It has been shown that the closed form of the LQU for quantum states in $\mathcal{H}^{2} \otimes \mathcal{H}^{d}$ is \cite{girolami2013}
\begin{equation}
\mathcal{U}_A=1-\lambda_{\rm max}(\mathcal{W}),
\end{equation}
where $\lambda_{\rm max}$ is the maximum eigenvalue of the $3 \times 3$ matrix $\mathcal{W}$ with elements $\mathcal{W}_{ij}={\rm Tr}\{\sqrt{\rho}(\sigma_{i}\otimes \mathbb{I})\sqrt{\rho}(\sigma_{j}\otimes \mathbb{I})\}$ and $\sigma_{i}$ $(i=1,2,3)$ represents the Pauli matrices. The closed form of the LQU for a large class of high-dimentional quantum states in $\mathcal{H}^{d_1} \otimes \mathcal{H}^{d_2}$ is \cite{wang2013}
\begin{eqnarray}
\label{CF}
\mathcal{U}_A=\frac{2}{d_1} - \lambda_{\rm max}(\mathcal{W}),
\end{eqnarray}
where $\mathcal{W}$ is a $(d_1^2-1)\times(d_1^2-1)$ matrix with elements
\begin{equation}
\label{W}
\mathcal{W}_{ij}=\text{Tr}\{\sqrt{\rho}(\lambda_{i}\otimes \mathbb{I}_{d_2})\sqrt{\rho}(\lambda_{j}\otimes \mathbb{I}_{d_2})\}-G_{ij} L,
\end{equation}
with
\begin{eqnarray}
&&G_{ij}=(g_{ij1}, \cdots , g_{ijk}, \cdots , g_{ijd_1^2-1}),\nonumber\\
&&L = (\text{Tr}(\rho \lambda_{1}\otimes \mathbb{I}_{d_2}), \cdots , \text{Tr}(\rho \lambda_{k}\otimes \mathbb{I}_{d_2}), \cdots , \text{Tr}(\rho \lambda_{d_1^2-1}\otimes \mathbb{I}_{d_2}))^T,\nonumber\\
&&  
\end{eqnarray}
and $\lambda_{j}$ $(j=1,\cdots,d^2-1)$ are the generators of SU($d$), namely,
\begin{equation}
\lambda_{j}=
\begin{cases}
\sqrt{\frac{2}{j(j+1)}}\left(\sum_{k\,=\,1}^{j}|k\rangle\langle k|-j|j+1\rangle\langle j+1|\right),j=1,...,d-1\\
|k\rangle\langle m|+|m\rangle\langle k| (1\leq k<m\leq d), j=d,...,\frac{d(d+1)}{2}-1\\
\mathrm{i}( |k\rangle\langle m|-|m\rangle\langle k|) (1\leq k<m\leq d), j=\frac{d(d+1)}{2},...,d^2-1\\
\end{cases},
\end{equation}
and $g_{ijk}=\frac{1}{4}\text{Tr}(\{\lambda_i,\lambda_j\}\lambda_k)$. It needs to be noted that the definition of the LQU requires $\Lambda^A$ not being degenerate, therefore the results after the simulation should be re-checked to make sure $\Lambda^A$ is non-degenerate when the LQU is maximized. This can be realized by the approach given in Ref. \cite{wang2013}. In the following, unless noted, the results are checked to be valid.

\subsection{Generalized amplitude damping}
For zero temperature environment, there only exists the transitions from higher energy levels to lower ones, in other words, the \emph{loss} of excitations. This kind of the transition is characterized by the amplitude damping (AD). In two dimensional case, the AD can be mathematically expressed by Kraus operators as
 \begin{eqnarray}
E_0&=&\left (\begin{array}{cc}
1& 0 \\
0& \sqrt{1-p}
\end{array}\right),
E_1=\left (\begin{array}{cc}
0& \sqrt{p} \\
0& 0
\end{array}\right),
\end{eqnarray}
where $p$ represents the transition probability from quantum state $|1\rangle$ to state $|0\rangle$. When the temperature of the environment is non-zero, the situation turns out to be more complicated since except for the the \emph{loss} of excitations, there exists the \emph{gain} of excitations. This process can be characterized by the generalized amplitude damping (GAD). Suppose the probability of losing the excitation $|1\rangle$ is $r$, then the probability of gaining the excitation is $1-r$. Therefore, in two dimensional quantum systems, the Kraus operators of the GAD are \cite{book}
\begin{eqnarray}
E_0&=&\sqrt{r}\left (\begin{array}{cc}
1& 0 \\
0& \sqrt{1-p}
\end{array}\right),
E_1=\sqrt{r}\left (\begin{array}{cc}
0& \sqrt{p} \\
0& 0
\end{array}\right),\nonumber\\
E_2&=&\sqrt{1-r}\left (\begin{array}{cc}
\sqrt{1-p}& 0 \\
0& 1
\end{array}\right),
E_3=\sqrt{1-r}\left (\begin{array}{cc}
0& 0 \\
\sqrt{p}& 0
\end{array}\right).
\label{GAD}
\end{eqnarray}
It needs to be noted that when $r=1$, the GAD reduces to the AD case.

For quantum systems consisting of three energy levels of $V$-configuration, the derivation of the Kraus operators of the GAD can be done naturally following the approach we have given. The results are
\begin{eqnarray}
&&E_0=\sqrt{r}\left (\begin{array}{ccc}
1& 0 &0 \\
0&\sqrt{1-p_1}& 0\\
0&0& \sqrt{1-p_2}\\
\end{array}\right),\nonumber\\
&&E_1=\sqrt{r}\left (\begin{array}{ccc}
0& \sqrt{p_1} &0 \\
0&0& 0\\
0&0& 0\\
\end{array}\right),
E_2=\sqrt{r}\left (\begin{array}{ccc}
0& 0 &\sqrt{p_2} \\
0&0& 0\\
0&0& 0\\
\end{array}\right),\nonumber\\
&&E_3=\sqrt{1-r}\left (\begin{array}{ccc}
\sqrt{1-p_1-p_2}& 0 &0 \\
0& 1& 0\\
0&0& 1\\
\end{array}\right),\nonumber\\
&&E_4=\sqrt{1-r}\left (\begin{array}{ccc}
0& 0 &0 \\
\sqrt{p_1}&0& 0\\
0&0& 0\\
\end{array}\right),
E_5=\sqrt{1-r}\left (\begin{array}{ccc}
0& 0 &0 \\
0&0& 0\\
\sqrt{p_2}&0& 0\\
\end{array}\right),\nonumber\\
&&
\label{GAD3}
\end{eqnarray}
where $p_1$ and $p_2$ are the transition probabilities from $|1\rangle$ and $|2\rangle$ to $|0\rangle$, respectively.

\begin{figure}[!htp]
\begin{centering}
\includegraphics[width=8cm]{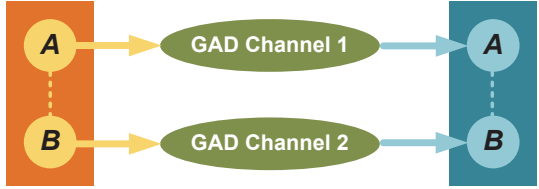}
\caption{(Color online) The two particles $A$ and $B$ undergo different GAD channels, i.e. 'GAD Channel 1' and 'GAD Channel 2' respectively.}
\label{fig2}
\end{centering}
\end{figure}

In our study, we let the two particles undergo different GAD channels as illustrated in Fig. \ref{fig2}. We assume the initial state is $\rho_i$, the state after decoherence is
\begin{equation}
\rho_f=\sum\limits_{i, j=0}^{n-1} (E_i\otimes E_j)\rho_i (E_i\otimes E_j)^{\dagger},
\end{equation}
where $n$ is the number of the Kraus operators.

\subsection{Weak measurement and measurement reversal}

\begin{figure}[!htp]
\begin{centering}
\includegraphics[width=9cm]{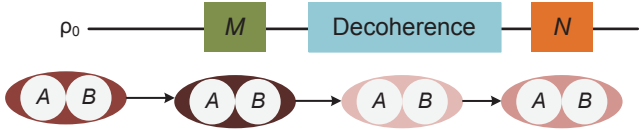}
\caption{(Color online) The procedure of enhancing quantum correlations under decoherence by weak measurement and measurement reversal. The intensity of the color in the ellipses stands for the quantum correlations between particles $A$ and $B$.}
\label{fig3}
\end{centering}
\end{figure}

The basic approach of enhancing quantum correlations by weak measurement and measurement reversal for two-partite quantum systems is illustrated in Fig. \ref{fig3}, where we call this scheme \emph{two-step enhancement of quantum correlations}. First, we apply weak measurement $M$ to the quantum system in order to push the initial state to a space with less decoherence effect. Then the two particles are put in the finite-temperature environment characterized by GAD channels. After decoherence, we apply the reversal measurement $N$ to recover the quantum correlations.

The weak measurement operator for qubit quantum systems in a general case is
\begin{eqnarray}
M^{(2)}
=\left (\begin{array}{cc}
1& 0 \\
0& m
\end{array}\right),
\end{eqnarray}
where $m \in [0, \infty)$, and when $0 \le m \le 1$, $M$ is a measurement partially collapsing the quantum system to the ground state $|0\rangle$, otherwise, $M$ partially collapses the quantum system to $|1\rangle$ \cite{wang2014}.
For qutrit quantum systems, the weak measurement is \cite{xiao2013}
\begin{eqnarray}
M^{(3)}
=\left (\begin{array}{ccc}
1& 0 & 0 \\
0& m^{(1)} &0\\
0 & 0 & m^{(2)}
\end{array}\right),\nonumber\\
&&
\end{eqnarray}
where $m_1, m_2 \in [0, \infty)$.

The measurement reversal operator for qubit and qutrit quantum systems are
\begin{eqnarray}
N^{(2)}
=\left (\begin{array}{cc}
n& 0 \\
0& 1
\end{array}\right),
N^{(3)}
=\left (\begin{array}{ccc}
n^{(1)} & 0 & 0 \\
0& n^{(2)} &0\\
0 & 0 & n^{(3)}
\end{array}\right),
\end{eqnarray}
where $n \in [0, \infty)$, and $0 \le n^{(1)}, n^{(2)}, n^{(3)} \le 1$.
It needs to be noted here that for qutrit quantum systems, we have constructed the operators $M^{(3)}$ and $N^{(3)}$ in more general forms than in Ref. \cite{xiao2013}.

\section{Enhancing quantum correlations}
\label{enhance}

\subsection{Qubit-qubit Bell state}

To demonstrate the approach, we will give our analysis in an explicit manner. The initial quantum state of the two particle is chosen as the qubit-qubit Bell state
\begin{equation}
\rho_0 = \frac{1}{2}(|00\rangle+|11\rangle)(\langle 00|+\langle 11|),
\end{equation}
and it is no surprise that $\mathcal{U}_A(\rho_0)=1$.

As we have stated above, the weak measurement is
\begin{eqnarray}
M&=&\left (\begin{array}{cc}
1& 0 \\
0& m_1
\end{array}\right)
\otimes 
\left (\begin{array}{cc}
1& 0 \\
0& m_2
\end{array}\right).
\label{WM}
\end{eqnarray}

After the weak measurement $M$ performed on $\rho_0$, the state becomes
\begin{equation}
\rho_1=\frac{M\rho_0 M^{\dagger}}{{\rm Tr}(\rho_0 M^{\dagger} M)}.
\end{equation}

Then we put the particles in a finite temperature environment to study the quantum correlations under decoherence. Without lose of generality, we choose $r_1=r_2=0.5$, $p_1=p_2=0.5$ in the GAD channels (Eq. (\ref{GAD})) and assume the two particles suffer from the same quantum noises (see Fig. \ref{fig2}). It can be calculated that without the weak measurement and measurement reversal, the LQU reduces to 0.134.

\begin{figure}[!htp]
\begin{centering}
\includegraphics[width=9cm]{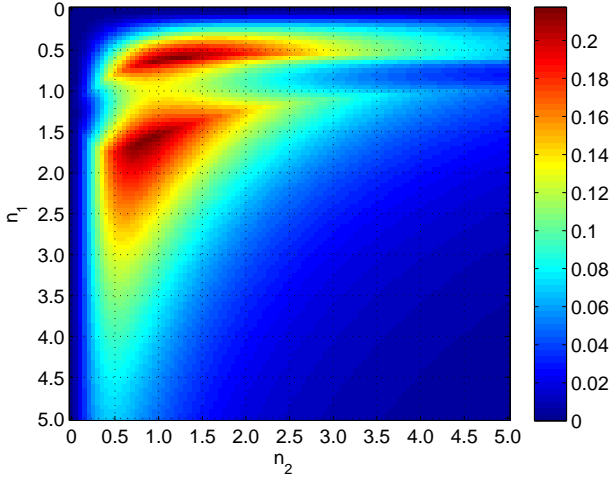}
\caption{(Color online) The LQU of $\rho_3$ versus $n_1$ and $n_2$ with fixed $m_1=1.285, m_2=0.760$.}
\label{fig4}
\end{centering}
\end{figure}

As the last step, we need to perform the measurement reversal
\begin{eqnarray}
N&=&\left (\begin{array}{cc}
n_1& 0 \\
0& 1
\end{array}\right)
\otimes 
\left (\begin{array}{cc}
n_2& 0 \\
0& 1
\end{array}\right),
\label{MR}
\end{eqnarray}
then the state is
\begin{equation}
\rho_3=\frac{N\rho_2 N^{\dagger}}{{\rm Tr}(\rho_2 N^{\dagger} N)}.
\end{equation}
Because of the large number of the parameters, we use the \emph{genetic algorithm} in our simulation. By optimizing upon $m_1$, $m_2$, $n_1$, $n_2$, the LQU of $\rho_3$ is maximized when $m_1=1.285, m_2=0.760, n_1=1.606, n_2=0.830$, in this case, $\mathcal{U}_A(\rho_1)=0.218$. The dependence of $\mathcal{U}_A(\rho_3)$ on $n_1$ and $n_2$ with fixed $m_1=1.285, m_2=0.760$ is shown in Fig. \ref{fig4}.

Therefore, we have shown that the weak measurements and measurement reversal have enhanced the quantum correlations between the particles under decoherence. 

\subsection{Non-symmetrical qubit-qubit mixed state}

We consider a general case in which the qubit-qubit state has no symmetry under the permutation of the two particles. The quantum state is chosen as
\begin{equation}
\rho_0=\frac{1}{2} |\psi\rangle\langle\psi|+\frac{1}{8}\mathbb{I},
\end{equation}
where $|\psi\rangle=\frac{1}{\sqrt{2}}|01\rangle+\frac{1}{2}|10\rangle+\frac{1}{2}|11\rangle$, and $\mathcal{U}_A(\rho_0)=0.096$.

\begin{figure}[!htp]
\begin{centering}
\includegraphics[width=9cm]{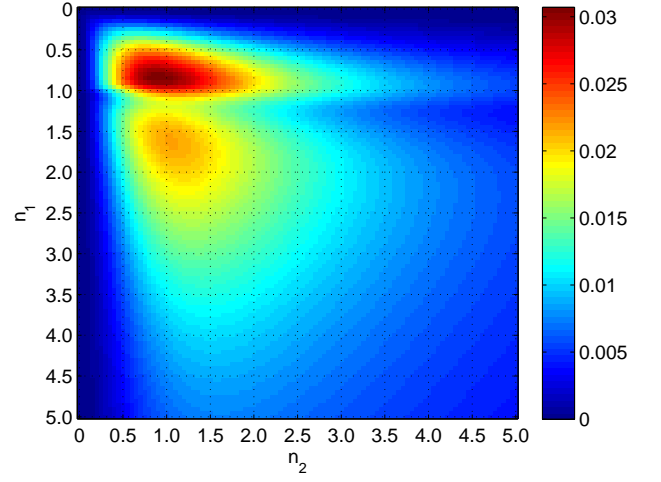}
\caption{(Color online) The LQU against $n_1$ and $n_2$ during the measurement reversal $N$ with fixed $m_1=1.65, m_2=1.20$.}
\label{fig5}
\end{centering}
\end{figure}

First we perform the weak measurement (see Eq. (\ref{WM})), then the two particles decoherence under different GAD channels where $r_1=r_2=0.5$, and $p_1=p_2=0.5$.
As the last step, the measurement reversal is performed. It can be optimized that the maximum of the LQU is given when $m_1=1.65, m_2=1.20, n_1=0.85, n_2=0.90$, and in this case, the LQU is 0.031. The LQU against $N$ with fixed $m_1=1.65, m_2=1.20$ is shown in Fig. \ref{fig5}.

Now we compare our results with and without weak measurement and measurement reversal. It can be easily calculated that if $M$ and $N$ are omitted, the decoherence of the two particles causes the quantum correlations drop rapidly. The LQU reduces to 0.019. We can see that the weak measurement and measurement reversal have enhanced the quantum system's ability against decoherence.

\subsection{Non-symmetrical qutrit-qutrit mixed state}
To illustrate our approach of enhancing the quantum correlations between qutrits, we consider
\begin{equation}
\rho_0 = \frac{1}{2} |\psi\rangle\langle\psi|+\frac{1}{18}\mathbb{I},
\end{equation}
where $|\psi\rangle=\frac{1}{2}|10\rangle+\frac{1}{\sqrt{2}}|02\rangle+\frac{1}{2}|21\rangle$, and $\mathcal{U}_A(\rho_0)=0.130$.

In this case, the weak measurement and measurement reversal operators are
\begin{eqnarray}
M&=&\left (\begin{array}{ccc}
1& 0 & 0 \\
0& m^{(1)}_1 &0\\
0 & 0 & m^{(2)}_1
\end{array}\right)
\otimes\left (\begin{array}{ccc}
1& 0 & 0 \\
0& m^{(1)}_2 &0\\
0 & 0 & m^{(2)}_2
\end{array}\right),\nonumber\\
N &=&\left (\begin{array}{ccc}
n^{(1)}_1& 0 & 0 \\
0& n^{(2)}_1 &0\\
0 & 0 & n^{(3)}_1
\end{array}\right)\otimes\left (\begin{array}{ccc}
n^{(1)}_2& 0 & 0 \\
0& n^{(2)}_2 &0\\
0 & 0 & n^{(3)}_2
\end{array}\right).
\end{eqnarray}
In this case, we consider a more general case in which $r=0.5, p_1=0.1, p_2=0.4$ in the GAD channels. As the weak measurement and measurement reversal operators are performed, and the LQU of the quantum state is maximized when $m^{(1)}_1=1.2745$, $m^{(2)}_1=1.29$, $m^{(1)}_2=1.1175$, $m^{(2)}_2=0.939$, $n^{(1)}_1=0.751$, $n^{(2)}_1=0.564$, $n^{(3)}_1=0.480$, $n^{(1)}_2=0.954$, $n^{(2)}_2=0.884$, $n^{(3)}_2=0.759$, where the LQU is 0.081. It need to be noted that without $M$ and $N$, the LQU after decoherence is 0.072.

To summarize, we list our results in Table. \ref{tab1}, where we have used '2D Bell', 'Non-symmetrical 2D', and '3D' to represent 'qubit-qubit Bell state', 'non-symmetrical qubit-qubit mixed state', and 'non-symmetrical qutrit-qutrit mixed state', respectively.

\begin{table}
\caption{The enhancement of the quantum correlations under decoherence by weak measurement and measurement reversal.}
\label{tab1}
\begin{tabular}{llll}
\noalign{\smallskip}\hline\noalign{\smallskip}\noalign{\smallskip}
(LQU of) Quantum states & 2D Bell& Non-symmetrical 2D & 3D\\
\noalign{\smallskip}\noalign{\smallskip}\hline\noalign{\smallskip}\noalign{\smallskip}
 Initial state & 1.0 & 0.096 & 0.130\\
Perform $M$ and $N$ & 0.218 & 0.031 & 0.081 \\
 Without $M$ and $N$ & 0.134 & 0.019 & 0.072 \\
\noalign{\smallskip}\hline
\end{tabular}
\end{table}

\section{Conclusion}
\label{con}

We use weak measurement and measurement reversal to enhance the quantum correlations in a quantum system consisting of two particles. The transitions of the quantum correlations of two- and three-dimensional quantum states during decoherence under generalized amplitude damping are investigated. We show that, after the weak measurement and measurement reversal, the joint system become robust against decoherence.

Except for the quantum correlations, we also care about the fidelity of the final output state. The fidelity of the final state $\rho_f$ is defined as \cite{jozsa1994}
\begin{equation}
F(\rho_i,\rho_f)=[{\rm Tr}\sqrt{\sqrt{\rho_f} \rho_i \sqrt{\rho_f}}]^2,
\end{equation}
where $\rho_i$ is the initial state.
In the first place we consider the case of the qubit-qubit Bell state. It can be calculated that after decoherence, the fidelity is reduced to 0.56. When $M$ and $N$ are performed, the fidelity of the final output is 0.52. As for the qutrit-qutrit mixed state, the fidelity is 0.925 and 0.964 with and without $M$ and $N$, respectively. 
However, the fidelity for the non-symmetrical qubit-qubit mixed state has been improved from 0.960 to 0.964 with weak measurement and measurement reversal.
To summarize, in most of the cases, due to the different physical meanings of the quantum correlations and fidelity, we can enhance the quantum correlations by sacrificing the fidelity \cite{wang2014}. But in some quantum states, it is still possible to improve (or keep) both the quantum correlations and the fidelity of the final state by using weak measurement and measurement reversal.

It needs to be noted that some of our examples can be implemented in nuclear magmatic resonance systems \cite{chuang2004}, linear photon systems \cite{moreva2006}, nitrogen-vacancy centres \cite{nv}, etc. We expect our work may find further theoretical and experimental applications.

\section*{Acknowledgements}
This work was supported by the National Natural Science Foundation of China under Grant Nos. 10874098 and 11175094 and the National Basic
Research Program of China under Grant Nos. 2009CB929402 and 2011CB9216002, GLL is a Member of Center of Atomic and Molecular
Nanosciences, Tsinghua University.

\end{document}